\documentclass[useAMS,usenatbib]{mn2e}
\usepackage{graphicx}
\usepackage{color}
\usepackage{amssymb,amsmath}
\usepackage{times}
\usepackage{psfig}

\def\spose#1{\hbox to 0pt{#1\hss}}
\newcommand\lsim{\mathrel{\spose{\lower 3pt\hbox{$\mathchar"218$}}
     \raise 2.0pt\hbox{$\mathchar"13C$}}}
\newcommand\gsim{\mathrel{\spose{\lower 3pt\hbox{$\mathchar"218$}}
     \raise 2.0pt\hbox{$\mathchar"13E$}}}

\title[A self-consistent interpretation of the GeV-TeV emission from  PKS 1424+240]{A self-consistent interpretation of the GeV-TeV emission from a distant blazar PKS 1424+240}
\author[Yan et al.]{Dahai Yan$^{1}$\thanks{E--mail: yandahai555@gmail.com}, Oleg Kalashev$^{2}$\thanks{E--mail: kalashev@inr.ac.ru}, Li Zhang$^{3}$\thanks{E--mail: lizhang@ynu.edu.cn}, Shuang-Nan Zhang$^1$\thanks{E--mail: zhangsn@ihep.ac.cn}\\
$^1$Key Laboratory of Particle Astrophysics, Institute of High Energy Physics,
Chinese Academy of Sciences, Beijing 100049, China\\
$^2$Institute for Nuclear Research of the Russian Academy of Sciences, Moscow 117312, Russia\\
$^3$Department of Physics, Yunnan University, Kunming 650091, China}

\begin{document}
\pagerange{\pageref{firstpage}--\pageref{lastpage}} \pubyear{2014}

\maketitle

\label{firstpage}

\begin{abstract}
we propose a scenario for self-consistent interpretation of GeV - TeV spectrum of a distant blazar PKS 1424+240. In this scenario, ultra-high energy (UHE) protons are assumed to exist in the blazar jet and produce gamma rays through synchrotron emission emitted by relativistic protons and pair cascades (resulted from $p\gamma$ interaction); meanwhile, some of these UHE protons may escape from the jet and are injected into intergalactic space. Therefore, we assume that UHE cosmic rays (CR) originate from relativistic protons in the jet and use energy-independent escape timescale to obtain UHECR injection spectrum. Both contributions of gamma rays injected by the source and secondary gamma rays produced in interactions of UHECRs emitted by the blazar with photon background during their propagation through intergalactic space are calculated. Our results show that this scenario is able to reproduce the GeV-TeV spectrum of PKS 1424+240 self-consistently in a broad range of redshifts $0.6<z<1.3$. The required relativistic jet power $L_p\simeq3\times10^{46}\ \rm erg\ s^{-1}$ only moderately depends on the assumed source redshift, while the proton escape timescale (and the injected UHECR luminosity) strongly depends on $z$. We compare the integral TeV fluxes predicted in our scenario with the sensitivity of the planned Cherenkov Telescope Array (CTA), and discuss the implications to future observations.
\end{abstract}


\section{Introduction}

In recent years, ground-based gamma-ray detectors have detected very high energy (VHE) gamma-ray radiations from distant blazars \citep[e.g.,][]{3c279VHE,Acciari2010}. In addition, recent analysis of the Large Area Telescope (LAT) on board the \emph{Fermi Gamma-ray Space Telescope} data show evidence of VHE gamma-ray emissions from 13 distant blazars with redshifts $z\sim0.6 - 1.5$ \citep{Neronov12}. Investigation of the VHE gamma rays from distant blazars is useful to constrain the density and evolution of the extragalactic
background light (EBL) \citep[e.g.,][]{3c279VHE}, to study the origin of UHECRs \citep[e.g.,][]{Takami13,essey14}, and to find evidence for hypothetical axion-like particles (ALPs) \citep[e.g.,][]{Meyer13}.

VHE gamma rays from distant blazars suffer serious EBL absorption. For instance, the attenuation factor of the flux at $\sim1\ $TeV emitted at $z=0.6$ due to EBL absorption is $\sim10^{-4}$ \citep{Dominguez11}. If VHE gamma rays from a blazar are produced in its jet, the intrinsic VHE spectrum after de-absorption would be very hard and not a simple power-law \citep[e.g.,][]{Archambault14}, which is hardly explained plausibly in leptonic models. In such a case, a lepto-hadronic jet model is used to explain VHE spectra of distant blazars \citep[e.g.,][]{bottcher09,Yan2014}, in which VHE emission is attributed to synchrotron emissions of relativistic protons and pair cascades created in proton-photon ($p\gamma$) interaction. However, the jet model can not explain the VHE emission from PKS 1424+240 if its redshift $z>0.7-0.8$ \citep{Yan2014}.
Alternatively, it is recently proposed that VHE gamma rays from distant blazars may be the secondary gamma rays produced in the rectilinear propagation of the UHECRs escaping from these blazars \citep[e.g.,][]{essey10a,essey10b}. In the latter case, UHECRs interact with background photons, i.e., EBL photons and microwave background (CMB) photons, which creates UHE electrons and photons through Bethe-Heitler (BH) pair production and photo-meson production. These UHE electrons and photons would interact with background photons again, and then pair cascades are induced; the secondary photons are inverse-Compton-scattered (ICS) CMB photons by the pair cascades. Because of the large mean interaction path of UHECRs, these secondary photons are produced relatively close to the Earth \citep[e.g.,][]{essey11,Lee,murase12}. The UHECR induced cascade model is successful in explaining the observed VHE spectra of extreme high-synchrotron-peaked BL Lacertae objects (HBLs) \citep[e.g.,][]{Aharonian2013,essey10a,essey14,murase12,Takami13}. In particular, UHECR induced cascade model is able to explain the VHE emission from a blazar with redshift $z>1$ \citep{Aharonian2013,essey14}. However, in the previous works the primary emission produced in the jet is either simply assumed \citep[e.g.,][]{essey14,Aharonian2013} or neglected \citep[e.g.,][]{Takami13}.

We would like to stress here that the secondary gamma-ray photons generated in the propagation of protons in intergalactic space may be related to the gamma-ray photons produced in the jet. In the lepto-hadronic jet model, UHE protons ($\gtrsim10^{18}$\ eV) exist in the jet \citep[e.g.,][]{Yan2014}, and some of these UHE protons may escape into intergalactic space, and then the secondary gamma-ray photons are produced. The escaping proton spectrum could be related to the relativistic proton spectrum in the jet via introducing an escape timescale \citep[e.g.,][]{Becker}. Therefore, the emissions produced in the jet and in the propagation of escaping protons can be evaluated simultaneously in one model.

PKS 1424+240 is a HBL. VHE gamma rays from this source have been detected by Imaging Atmospheric Cherenkov Telescopes (IACTs) \citep{Acciari2010,Aleks2014,Archambault14}. Its redshift is uncertain, but a firm lower limit of $z>0.6$ is determined \citep{Furniss13}. A photometric redshift upper limit of $z\leq1.10$ was reported by \citet{rau}.
Based on its gamma-ray spectra, the redshift upper limits of $z\leq0.81$, $z\leq1.00$ and $z\leq1.19$ were reported by \citet{Aleks2014}, \citet{Scully} and \citet{yang}, respectively. The upper limit of $z\leq1.00$ \citep{Scully} is EBL-model-independent.
In our previous paper \citep{Yan2014}, a lepto-hadronic jet model was used to explain the VERITAS spectrum in 2009 reported in \citet{Archambault14}. It was shown that this VHE spectrum could be explained in the jet model with a very low EBL density if its redshift $0.6<z<0.75$. \citet{essey14} presented that the UHECR induced cascade model could reproduce the VHE spectrum reported in \citet{Acciari2010} with the redshift up to 1.3.

In this work, we propose a scenario for explaining GeV - TeV emission from a distant balzar, where both of primary gamma rays produced in the jet and secondary gamma rays created in UHECRs propagation are considered simultaneously (hereafter called jet+UHECR model). We apply this scenario to study the formation of the VERITAS spectrum of PKS 1424+240 observed in 2009 and reported in \citet{Archambault14}. In the jet+UHECR model, the GeV-TeV spectrum is explained self-consistently and energy budget is discussed in detail.
We use the cosmology parameters $H_{0}=70\rm \ km\ s^{-1}\ Mpc^{-1}$, $\Omega_{\rm m}=0.3$ and $\Omega_{\rm \Lambda}=0.7$ in the calculations.

\section{The model}

In the jet+UHECR model, we assume high energy electrons and UHE protons coexist in emission region of radius $R^{\prime}_{b}$ containing a magnetic field of strength $B$ in the blazar jet. Because of the beaming effect, particles and photons from this region are strongly boosted, and the Doppler factor is $\delta_{\rm D}$.
The details of high energy emission processes in the jet are described in lepto-hadronic jet models, \citep[e.g.,][]{bottcher13,Cerruti,dimit2014,Yan2014}. In such a model, the following six processes are included: (1) synchrotron and SSC emissions of primary electrons; (2) proton-photon pion production; (3) proton synchrotron emission; (4) BH pair production; (5) photon-photon pair production; and (6) synchrotron emission of UHE-photons-induced pair cascades.

Some UHE protons in the jet may escape from the source and propagate in the intergalactic space. The injected (into intergalactic space) proton spectrum is assumed to be the escaping proton spectrum from the emission region in the jet. The injected proton rate, $Q(\gamma_{p})$, at energy $\gamma_{p}$ in the host galaxy frame can be expressed as \citep[e.g.,][]{Dermerbook}
\begin{equation}
\gamma^2_{p}Q(\gamma_{p})=\frac{\delta_{\rm D}^{4}\gamma^{\prime 2}_{p}N^{\prime}(\gamma^{\prime}_{p})}{t_{\rm esc}},
\label{inj}
\end{equation}
where $N^{\prime}(\gamma^{\prime}_{p})$ is the emitting proton distribution at energy $\gamma^{\prime}_{p}$ in the emission region frame, and $\gamma_{p}=\delta_{\rm D}\gamma^{\prime}_{p}$. The escape timescale, $t_{\rm esc}$, is written as $t_{\rm esc}=\eta\frac{R^{\prime}_{\rm b}}{c}$, and $\eta>$ is a constant. The escape process does not affect the gamma-ray spectrum produced in the jet \citep{Dermerbook,bottcher13}.
Previous studies have shown that the injected proton spectrum has little effect on the shape of secondary photon spectrum formed in the propagation of cosmic rays, and it affects only the normalization of the photon spectrum \citep[e.g.,][]{essey11}. Our calculations show that if the redshift of the source is large (e.g., $z>0.6$), the secondary VHE spectrum around $\sim1\ $TeV is affected by the maximum energy of injected protons ($E_{p, \rm max}$) (also see Fig. 2 in Aharonian et al. 2013).

The details of the interaction processes in the UHECRs (assumed to be protons here) propagation can be found in \citet{Lee} and \citet{Bhattacharjee}, which include the above processes (2), (4), (5) and double pair production, and IC scattering of CMB by pair cascades. We use the code TransportCR developed by Kalashev et al. \citep{Arisaka,Gelmini12,Kalashev14}, which is used in a number of published papers \citep[e.g.,][]{Arisaka,Kalashev09,Gelmini12,Kalashev13}, to calculate the secondary emission produced in UHECRs propagation. The code simulates the standard interaction processes in UHECRs propagation by solving transport equation in one dimension (see details in Kalashev \& Kido 2014).
Secondary gamma rays can only reach us from the direction of the source if deflection of protons and electron-photon cascades by intergalactic magnetic field (IGMF) are small compared to the angular resolution of the instrument.
The corresponding deflection angle for protons is \citep[e.g.,][]{essey10a}
\begin{equation}
\theta_{p}\sim 0.1^{\circ}\left(\frac{B_{\rm IGMF}}{10^{-14}\ \rm G}\right)\left(\frac{4\times10^{16}\ \rm eV}{E_{p}}\right)\left(\frac{D}{\rm Gpc}\right)^{1/2}\left(\frac{l_{\lambda}}{\rm Mpc}\right)^{1/2}.
\end{equation}
Here, $B_{\rm IGMF}$ is the intergalactic magnetic field (IGMF) strength and $l_{\lambda}$ is the coherence length of magnetic field; $D$ is the distance to the source and $E_{p}$ is the proton energy. At present the intergalactic magnetic field (IGMF) is very poorly known. The theoretical and observational constraints on the mean
IGMF strength $B_{\rm IGMF}$ and correlation length $l_{\lambda}$ are summarized in the review by ~\citet{Durrer}:
$$10^{-18}\ \rm G \lsim \ \  B_{\rm IGMF} \ \  \lsim 10^{-9}\ \rm G, $$
$$l_{\lambda} \ \ \gsim \ \ 1 \rm pc. $$
The lower limits on $B_{\rm IGMF}$ derived from GeV-TeV spectra of blazars vary from $10^{-18}$ G to $10^{-15}$ G \citep[e.g.,][]{Neronov10,Tavecchio10,Dermer11}. Taking into account the secondary contributions from cosmic rays to blazars spectra, the 95\% confidence interval for $B_{\rm IGMF}$ of $(0.01-30)\times10^{-15}\ $G was reported by \citet{essey11b}. The observed evidences of GeV pair halos around blazars indicate $B_{\rm IGMF}$ in the range $(0.01-1)\times10^{-15}\ $G \citep[e.g.,][]{Ando,Chen}.
In the following, we adopt $B_{\rm IGMF}=10^{-15}\ $G and $l_{\lambda}=1\ $Mpc (unless stated otherwise).
For such a field the deflection angle of protons with energy $E_{p}\gtrsim10^{17}\ $eV emitted by a distant blazar is smaller than the angular resolution of IACTs and \emph{Fermi}-LAT.

The deflection of pair cascades can be estimated assuming that the magnetic field correlation
length is always higher than the electrons mean free path. In TransportCR code we mimic the deflections by assigning a finite lifetime to the cascade electrons. While this is a simplification, the results were tested against the full Monte-Carlo calculations and they agree sufficiently well for our purposes.

In our preliminary calculations, we see that the deflection of pair cascades mainly affects secondary gamma rays below $\sim$30\ GeV assuming $B_{\rm IGMF}\sim10^{-15}-10^{-14}\ $G and $l_{\lambda}=1\ $Mpc.
Containment radius of \emph{Fermi}-LAT PSF strongly depends on energy below $\sim5\ $GeV \citep{Ackermann13}. Practically, we simply adopt the PSF of $0.5^\circ$, the 95\% containment radius of the on-orbit PSF for front events at $\sim5\ $GeV \citep{Ackermann13}. We will discuss the effect of this treatment on our results in Section~\ref{DC}.

\section{results}
\label{res}

\begin{figure}
  \begin{center}
  \begin{tabular}{c}
\hspace{-0.90cm}
     \includegraphics[width=100mm,height=80mm]{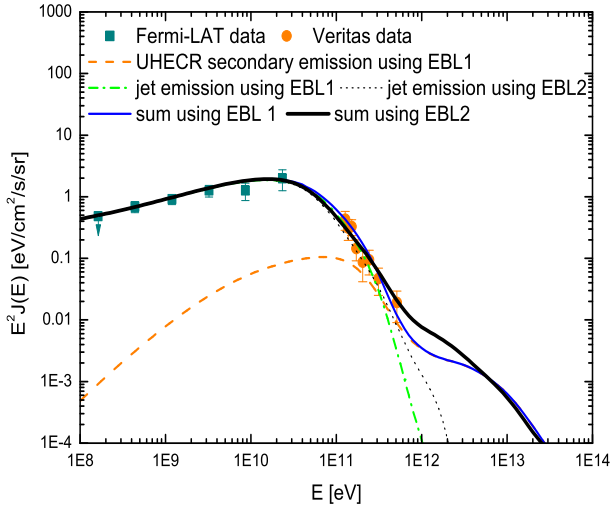}
\end{tabular}
  \end{center}
\caption{Reproduction of the observed gamma-ray spectra of PKS 1424+240 in the jet+UHECR model with $z=0.6$.
EBL1 refers to the EBL model of \citet{Franceschini}, and EBL2 refers to the EBL model of \citet{Inoue13}.
 The details of the observed data can be found in \citet{Archambault14}.} \label{z06}
\end{figure}

\begin{figure}
  \begin{center}
  \begin{tabular}{c}
\hspace{-0.90cm}
     \includegraphics[width=90mm,height=80mm]{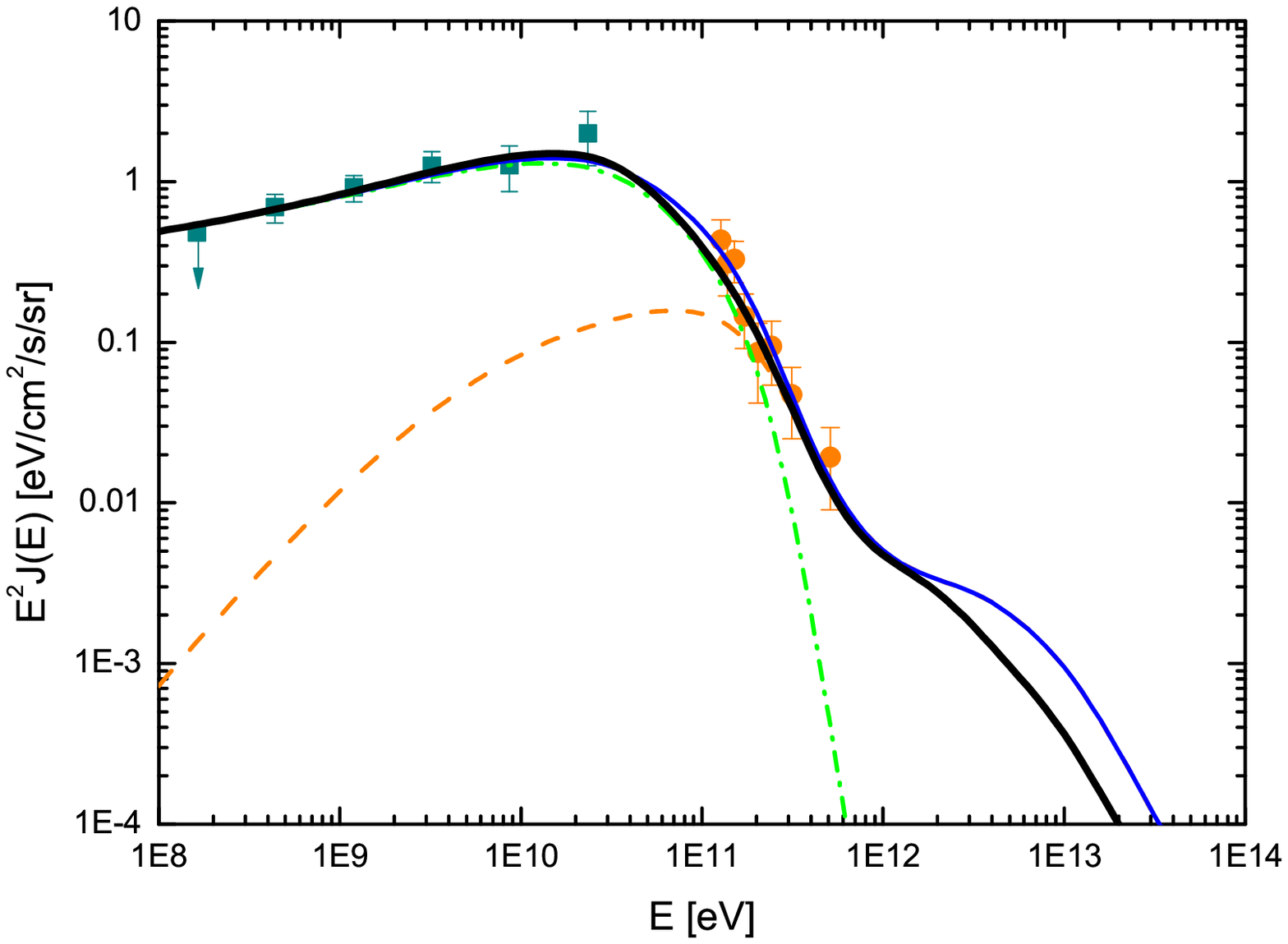}
\end{tabular}
  \end{center}
\caption{Same as Fig.~\ref{z06} but with $z=0.8$. } \label{z08}
\end{figure}

\begin{figure}
  \begin{center}
  \begin{tabular}{c}
\hspace{-0.90cm}
     \includegraphics[width=90mm,height=80mm]{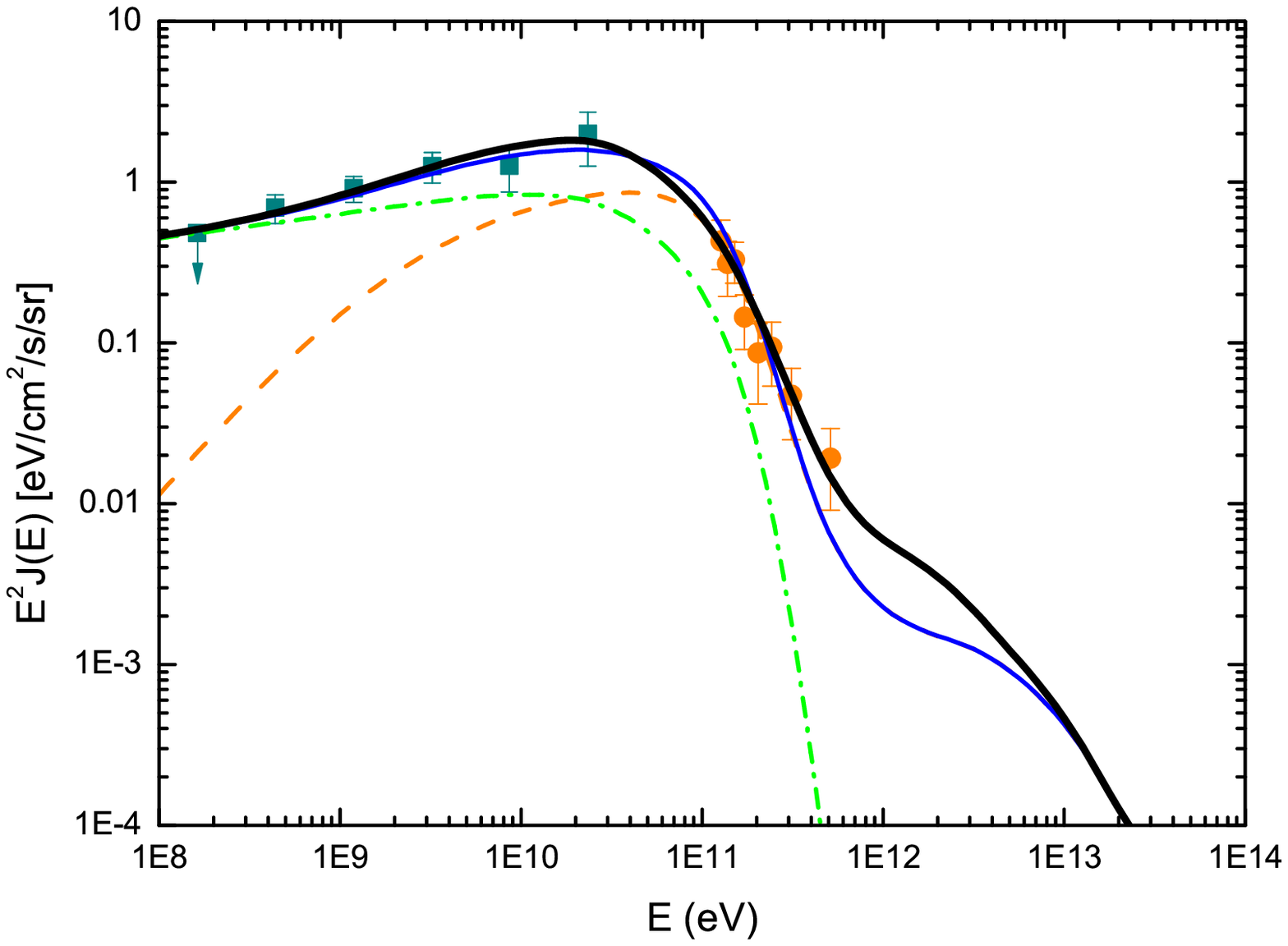}
\end{tabular}
  \end{center}
\caption{Same as Fig.~\ref{z06} but with $z=1.0$.} \label{z10}
\end{figure}

\begin{figure}
  \begin{center}
  \begin{tabular}{c}
\hspace{-0.90cm}
     \includegraphics[width=90mm,height=80mm]{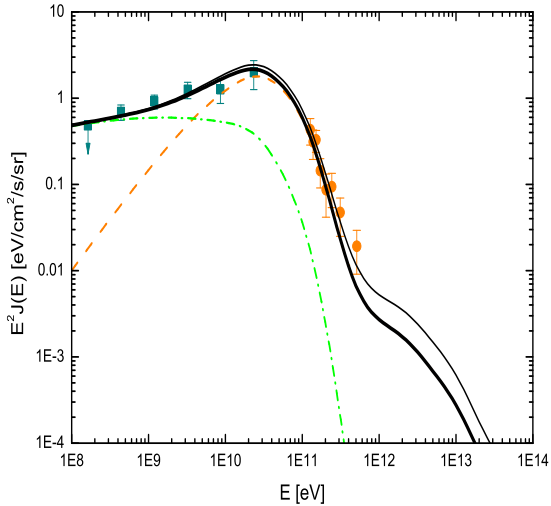}
\end{tabular}
  \end{center}
\caption{Same as Fig.~\ref{z06} but with $z=1.3$ and $B_{\rm IGMF}=6\times10^{-15}\ $G, and only results using EBL of \citet{Inoue13} are showed. The thin-solid line presents the total flux using the maximum energy of injected protons of $7\times10^{18}\ $eV.} \label{z13}
\end{figure}

\begin{figure}
  \begin{center}
  \begin{tabular}{c}
\hspace{-0.90cm}
     \includegraphics[width=90mm,height=80mm]{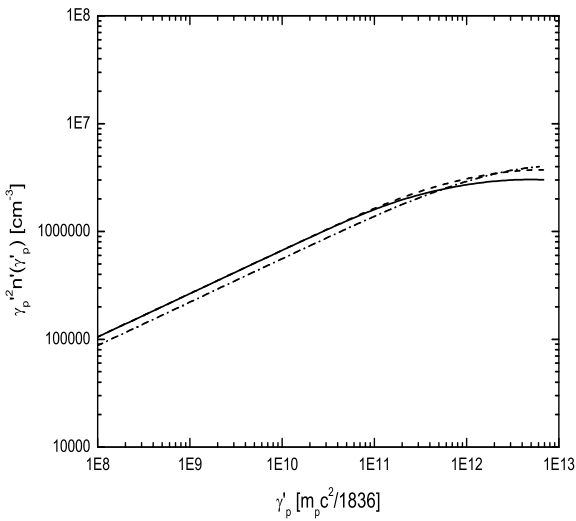}
\end{tabular}
  \end{center}
\caption{Spectral density of emitting protons in the jet in the emission region frame in the cases of $z=0.6$ (dash-dotted line), $z=0.8$ (dashed line) and $z=1.0$ (solid line).} \label{pde}
\end{figure}

The model parameters $B$, $R^{\prime}_{\rm b}$ and $\delta_{\rm D}$ can be derived through modeling the low-energy component of SED. The values derived in \citet{Yan2014} are used here, i.e., $B=15\ $G, $R^{\prime}_{\rm b}=4.7\times10^{15}\ $cm, and $\delta_{\rm D}=30$. We can make the primary-electron synchrotron emission to match the observed low-energy component of SED through adjusting the parameters of electron distribution when using different redshifts. The maximum energy of protons of $E_{p, \rm max}\approx10^{20}\ $eV in the host galaxy frame derived in \citet{Yan2014} is also used.
We here focus on the reproduction of the high energy SED in the jet+UHECR model.

Figs.~\ref{z06}-\ref{z13} show the reproduction of the observed GeV-TeV spectrum in the jet+UHECR model using EBL models of \citet{Franceschini} and \citet{Inoue13}, and different redshifts.

In the case of $z=0.6$ (Fig.~\ref{z06}), it can be seen that the contribution of the secondary gamma rays produced in the UHECRs propagation to the SED is suppressed by using a large $\eta\approx480$ because the jet emission is not yet substantially absorbed. The power of relativistic protons in the jet is $L_p\approx2.4\times10^{46}\ \rm erg\ s^{-1}$. The corresponding injected (escaping) proton power is $L_{\rm inj}\approx5.0\times10^{43}\ \rm erg\ s^{-1}$. It is noted that if a lager $L_{\rm inj}$ (e.g., $L_{\rm inj}\gtrsim8.0\times10^{43}\ \rm erg\ s^{-1}$) is used, secondary gamma rays will exceed the observed VHE spectrum above $\sim0.2\ $TeV. The two EBL models make the spectra around 1\ TeV different, and the total flux at $\sim1$TeV obtained assuming EBL model of \citet{Inoue13} is higher than that assuming EBL model of \citet{Franceschini}.

In the case of $z=0.8$ (Fig.~\ref{z08}), the contribution of secondary emission is improved to match the VHE observation, which becomes dominant above $\sim0.2\ $TeV.  We use $L_p\approx3.2\times10^{46}\ \rm erg\ s^{-1}$ and $\eta=190$, resulting in $L_{\rm inj}\approx1.7\times10^{44}\ \rm erg\ s^{-1}$. In this case, the TeV spectra calculated by using different EBL models are similar.

In the case of $z=1.0$ (Fig.~\ref{z10}), the GeV-TeV spectrum also can be reproduced in the jet+UHECR model.
To match the observed VHE spectrum, the secondary component is substantially enhanced, which becomes dominant above $\sim20\ $GeV. Due to the large redshift, the optical--X-ray luminosity also becomes large, which leads to a high $p\gamma$ interaction rate and SSC emission. Therefore, to avoid the gamma-ray emission from the jet exceeding the observed GeV spectrum, the same $L_p$ as in the case of $z=0.8$ is used. However, different values of $\eta$ are needed for different EBL models, i.e., $\eta=70$ for EBL model of \citet{Inoue13} and $\eta=47$ for EBL model of \citet{Franceschini}. The corresponding injected proton power is $L_{\rm inj}\approx4.4\times10^{44}\ \rm erg\ s^{-1}$ and $L_{\rm inj}\approx7.1\times10^{44}\ \rm erg\ s^{-1}$, respectively. In this case, it seems that the spectrum calculated by using EBL model of \citet{Inoue13} gives a better reproduction of the observed TeV spectrum than using EBL model of \citet{Franceschini}.

In the case of $z=1.3$ (Fig.~\ref{z13}), we only show the results calculated by using EBL model of \citet{Inoue13}.
It can be seen that the secondary component becomes dominant above several GeV. $L_p\approx2.9\times10^{46}\ \rm erg\ s^{-1}$ is used, and a smaller $\eta=10$ and a larger $B_{\rm IGMF}=6\times10^{-15}\ $G are needed. The corresponding injected proton power is $L_{\rm inj}\approx2.9\times10^{45}\ \rm erg\ s^{-1}$, which is ten percent of $L_p$.
In Fig.~\ref{z13}, we also show the effect of maximum energy of injected protons on the secondary photon spectrum.
The maximum energy of injected protons of $10^{20}\ $eV and $7\times10^{18}\ $eV (in the host galaxy frame) are used to calculate the results presented by the thick and thin solid lines in Fig.~\ref{z13}, respectively. Using a smaller $E_{p, \rm max}$, a better reproduction of the observed TeV spectrum is obtained. A smaller $\eta\approx3$ is required in the case of $E_{p, \rm max}\approx7\times10^{18}\ $eV.

One can see that the jet+UHECR model is able to reproduce the GeV-TeV spectrum of PKS 1424+240 assuming its redshift in the range of 0.6 -- 1.3. The relativistic proton power in the jet which is constrained by the \emph{Fermi}-LAT spectrum is insensitive to the redshift. However, for a higher redshift, a larger injected proton power is needed; for instance from $z=0.6$ to 1.3 the required $L_{\rm inj}$ increases from $5.0\times10^{43}\ \rm erg\ s^{-1}$ to $2.9\times10^{45}\ \rm erg\ s^{-1}$ (or $10^{46}\ \rm erg\ s^{-1}$, depending on $E_{p, \rm max}$). The effect of IGMF on modeling of GeV-TeV SED is negligible when $z<1$. When $z\gtrsim1$, the contribution of secondary emission to the SED is dominant over primary emission above  $\sim20\ $GeV, and the modeling result becomes sensitive to $B_{\rm IGMF}$.

In Fig.~\ref{pde}, we show the relativistic proton spectral densities in the emission region frame providing the jet emissions in Figs.~\ref{z06}-\ref{z10}, and $n^{\prime}(\gamma^{\prime}_p)=N^{\prime}(\gamma^{\prime}_p)/(4/3\pi R^{\prime 3}_{\rm b})$. There are no peaks in these distributions because the energy loss rates of relativistic
protons are dominated by adiabatic losses \citep[e.g.,][]{Yan2014,Cerruti}.
The corresponding injected proton spectra are calculated by using Equation~(\ref{inj}). The lowest energy of protons $E_{p,\rm min}$ does not affect the secondary emissions. Using $E_{p,\rm min}=10\ $TeV and $0.1\ $EeV, we derive the same results. The relevant proton power is insensitive to $E_{p,\rm min}$ because the spectral index is smaller than 2, is very hard but still possible in certain acceleration scenarios \citep[e.g.,][]{VV05,SB}.

\section{conclusion and discussion}
\label{DC}

As demonstrated above, the jet+UHECR model can reproduce the observed VHE spectrum of PKS 1424+240 in a broader range of redshifts, from the observational lower limit 0.6 to 1.3, than the jet model proposed by \citet{Yan2014}, thanks to the secondary component produced in the UHECRs propagation. Our modeling results are roughly consistent with the results showed in \citet{essey14}.
The advantages of the jet+UHECR model are that
it can explain the GeV-TeV spectrum of PKS 1424+240 self-consistently and it clearly shows that the jet can provide energy to produce the secondary gamma-ray emission.
Indeed, as long as a small fraction of the relativistic proton power in the jet is injected into intergalactic space, the produced secondary gamma-ray photons can account for the observed VHE spectrum from a blazar residing at a redshift of $\sim1$. The required relativistic proton power of $L_p\sim10^{46}\ \rm erg\ s^{-1}$ (dominating the jet power) can be provided by a $10^9\ M_\odot$ black hole with the Eddington luminosity of $L_{\rm Edd}\sim10^{47}\ \rm erg\ s^{-1}$.  The isotropic equivalent luminosity $L_{\rm iso}$ can be much greater than the absolute power $L$ (e.g., $L_p$ and $L_{\rm inj}$ ): $L_{\rm iso}=L/f_b$, where $f_b\approx(1-\rm cos\theta_{\rm jet})/2\approx\theta^2_{\rm jet}/4\approx1/4\delta_{\rm D}^2$ \citep{essey11}. Taking $\delta_{\rm D}=30$, we derive $L_{p,\rm iso}\sim10^{49}\ \rm erg\ s^{-1}$. The absolute injected proton power of $L_{\rm inj}\sim10^{44-45}\ \rm erg\ s^{-1}$ (the corresponding isotropic equivalent luminosity $L_{\rm inj,iso}\sim10^{46-48}\ \rm erg\ s^{-1}$) is larger than that of lower redshift blazars derived in \citet{essey11}. It is clear that $L_{\rm inj}$ and $L_{\rm inj,iso}$ strongly depend on the assumed redshift,
and $L_{\rm inj,iso}$ is the upper limit of cosmic-ray luminosity emitted by PKS 1424+240. The dependence of $L_{\rm inj,iso}$ on the redshift may effect the estimates of the contribution from blazars to the observed cosmic rays and the blazar number emitting TeV flux above a certain flux limit.

We show the effect of $E_{p, \rm max}$ on the secondary TeV spectrum in the case of $z=1.3$.
 Using a smaller $E_{p, \rm max}\approx8\times10^{18}\ $eV, the modeling result becomes better; however, the required larger $L_{\rm inj}\approx10^{46}\ \rm erg\ s^{-1}$ is about 30\% of $L_p$.
 When the redshift approaches to unity, the contribution from secondary emission becomes dominant above $\sim20\ $GeV. This means that in the frame of the jet+UHECR model, if the redshift of PKS 1424+240 is $\sim1$, no variability on timescale of months or shorter should be detected above $\sim20\ $GeV. Therefore, GeV variability features can be used to test the jet+UHECR model and constrain redshift of PKS 1424+240. The escape process of particles is assumed to be energy-independent in this work. In an energy-dependent escape scenario, the maximum energy of escaping particles is limited by the escape timescale \citep[e.g.,][]{Dermer14}.

 Let us discuss the effect of the simple treatment of the \emph{Fermi}-LAT PSF. Because of the larger realistic PSF angle below $\sim5\ $GeV than $0.5^\circ$, the secondary gamma-ray flux below $\sim5\ $GeV was underestimated, which can only effect the results using $z\gtrsim1$ where the contribution of the secondary gamma-ray flux to the observed GeV spectrum can not be neglected. However, the secondary gamma-ray flux below $\sim5\ $GeV is also affected by the value of $B_{\rm IGMF}$. If the exact PSF is used, in order to reproduce the GeV-TeV spectrum, a larger value of $B_{\rm IGMF}$ ($\sim10^{-14}\ $G) than that adopted in cases of $z=1.0$ and $1.3$ will be needed to suppress secondary gamma-ray flux below $\sim5\ $GeV.

 \begin{figure}
  \begin{center}
  \begin{tabular}{c}
\hspace{-0.90cm}
     \includegraphics[width=90mm,height=80mm]{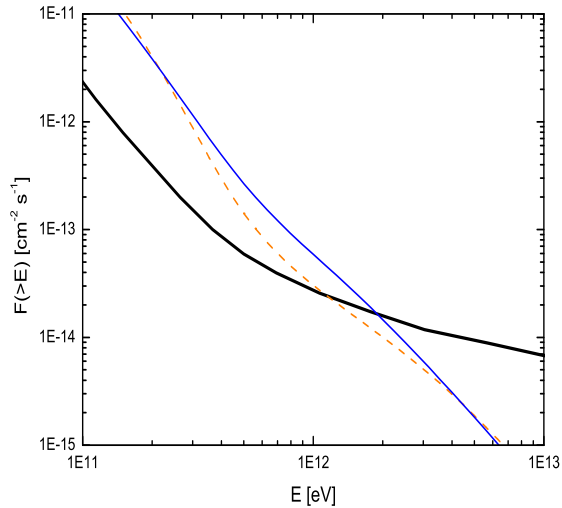}
\end{tabular}
  \end{center}
\caption{ Integral total fluxes in the case of $z=0.6$ using EBL models of \citet{Franceschini} (dashed line) and \citet{Inoue13} (solid line). The thick-solid line is the integral
sensitivity goal of Cherenkov Telescope Array (CTA) for a 50 hours observation \citep{Actis}.} \label{CTA}
\end{figure}

VHE emission from distant blazars may have profound impact on several areas related to gamma-ray astronomy \citep{essey12}, e.g., the measurement of EBL \citep[e.g.,][]{3c279VHE}, the origin of UHECRs \citep[e.g.,][]{Takami13} , and the finding of ALPs \citep[e.g.,][]{Meyer14}. If the secondary emission produced in the UHECRs propagation indeed contributes to the observed VHE emission of blazars, the high density EBL model of \citet{stecker06} could also be favored \citep{essey11}. Next, let us discuss the implications of our results. In Fig.~\ref{CTA}, we show the comparison with the integral fluxes predicted by our jet+UHECR model in the case of $z=0.6$ and the integral sensitivity goal of the planned project CTA. One can see that it is likely for CTA to detect VHE emission $\sim$1\ TeV from PKS 1424+240. \citet{Inoue14} showed that a large number of blazars can be detected by CTA if the secondary gamma rays are taken into account and the future survey of CTA will provide evidence of the secondary gamma-ray scenarios.
Our model predicts that no signals above several TeV from PKS 1424+240 can be detected by CTA even if the EBL density is as low as that estimated by \citet{Inoue13}. If the signals above several TeV from PKS 1424+240 are detected in the future, then an alternative way is needed to explain it, e.g, the existence of ALPs. A firm determination of the redshift of PKS 1424+240 will put strong constraints on the relevant issues.

\section*{Acknowledgments}
 We thank the referee for his/her valuable comments. This work is partially supported by the National Natural Science Foundation of China (NSFC 11433004). O.K. was supported by Russian Science Foundation grant 14-12-01340. SNZ acknowledges partial funding support by 973 Program of China under grant 2014CB845802, by the National Natural Science Foundation of China (NSFC) under grant Nos. 11133002 and 11373036, by the Qianren start-up grant 292012312D1117210, and by the Strategic Priority Research Program ``The Emergence of Cosmological Structures'' of the Chinese Academy of Sciences (CAS) under grant No. XDB09000000.

\bibliography{refernces}

\end{document}